\documentclass[preprint,amsmath,amssymb,superscriptaddress,longbibliography,aps,floatfix,footinbib]{revtex4-2}
\usepackage[]{graphicx}
\usepackage{tabularx}
\usepackage[usenames,dvipsnames]{color}
\usepackage{soul}
\usepackage{bm}
\usepackage{amsmath}
\usepackage{lineno}
\usepackage{gensymb}
\usepackage[utf8]{inputenc}

\usepackage{amsfonts}
\usepackage{mathrsfs}
\usepackage{graphicx}
\usepackage{dcolumn}
\usepackage{bm}
\usepackage{color}

\usepackage[colorlinks,bookmarks=false,citecolor=blue,linkcolor=red,urlcolor=blue]{hyperref}
\usepackage{multirow}
\usepackage{physics}
\usepackage{siunitx}
\DeclareSIUnit{\barpressure}{bar}
\DeclareSIUnit\angstrom{\protect \text {Å}}

\begin{document}

\title{Discovery of magnetic-field-tunable density waves in a layered altermagnet}

\author{Christopher Candelora}
\affiliation{Department of Physics, Boston College, Chestnut Hill, Massachusetts 02467, USA}

\author{Muxian Xu}
\affiliation{Department of Physics, Boston College, Chestnut Hill, Massachusetts 02467, USA}

\author{Siyu Cheng}
\affiliation{Department of Physics, Boston College, Chestnut Hill, Massachusetts 02467, USA}

\author{Alessandro De Vita}
\affiliation{Institut für Optik und Atomare Physik, Technische Universität Berlin, Straße des 17 Juni 135, 10623 Berlin, Germany}
\affiliation{Fritz Haber Institut der Max Planck Gesellshaft, Faradayweg 4--6, 14195 Berlin, Germany}

\author{Davide Romanin}
\affiliation{Université Paris-Saclay, CNRS, Centre de Nanosciences et de Nanotechnologies, 91120, Palaiseau, Paris, France}

\author{Chiara Bigi}
\affiliation{Synchrotron SOLEIL, F-91190 Saint-Aubin, France}

\author{My Bang Petersen}
\affiliation{Department of Physics and Astronomy, Interdisciplinary Nanoscience Center, Aarhus University, 8000 Aarhus C, Denmark}

\author{Alexander LaFleur}
\affiliation{Department of Physics, Boston College, Chestnut Hill, Massachusetts 02467, USA}

\author{Matteo Calandra}
\affiliation{Department of Physics, University of Trento, Via Sommarive 14, Povo 38123, Italy}

\author{Jill Miwa}
\affiliation{Department of Physics and Astronomy, Interdisciplinary Nanoscience Center, Aarhus University, 8000 Aarhus C, Denmark}

\author{Younghun Hwang}
\affiliation{Electricity and Electronics and Semiconductor Applications, Ulsan College, Ulsan 44610, Republic of Korea}

\author{Ziqiang Wang}
\affiliation{Department of Physics, Boston College, Chestnut Hill, Massachusetts 02467, USA}

\author{Federico Mazzola}
\affiliation{CNR-SPIN, c/o Complesso di Monte S. Angelo, IT-80126 Napoli, Italy}

\author{Ilija Zeljkovic}
\affiliation{Department of Physics, Boston College, Chestnut Hill, Massachusetts 02467, USA}

\maketitle

\section{Abstract}
\textbf{Altermagnets recently came into the spotlight as a new class of magnetic materials, arising as a consequence of specific crystal symmetries \cite{Hayami2019, Smejkal2020, Yuan2020, Mazin2021}. They are characterized by a spin-polarized electronic band structure similar to ferromagnets, but with net zero magnetization, and touted as a promising platform to host a slew of exotic properties, many of which are yet to be explored \cite{Smejkal2022_2, Parthenios2025DWtheory, Sim2024PDW}. Here we study a new layered triangular lattice altermagnet, Co-intercalated NbSe$_2$ using scanning tunneling microscopy and spectroscopy (STM/S). Differential conductance d$I$/d$V$ spectra at low temperature reveal a surprising partial gap opening centered at the Fermi level, which is not captured by density functional theory calculations of the system in the pure altermagnetic state. Spatial mapping using spectroscopic-imaging STM and spin-polarized STM further reveals emergent tri-directional charge and spin density modulations with a 2$a_0$ wave length. Interestingly, we discover that out-of-plane magnetic field can serve as a knob to tune the amplitudes of the modulations as well as alter the overall electronic density-of-states in a manner that is strongly dependent on the field direction and strength. This can be attributed to the tilting of spins by the external magnetic field, which can have profound implications on the electronic properties of the altermagnet. By providing elusive atomic-scale insights, our work uncovers a tunable density wave accompanied by concomitant changes in the electronic band structure, and sets the foundation for studies of correlated electronic phenomena in altermagnets.}



\section{Introduction}
For decades, the field of magnetism has focused on classifying collinear magnetic orders into two main categories: antiferromagnetic orders or ferromagnetic orders. The former is a crystal-symmetry compensated state with a net zero magnetization and Kramers degenerate bands, while the latter arises from spin splitting of the electronic band structure, generating a net magnetization. In recent years, several groups reported theoretical predictions of collinear antiferromagnets demonstrating various time-reversal symmetry breaking phenomena and spin-polarized behaviors \cite{Hayami2019, Smejkal2020, Yuan2020, Mazin2021}, subsequently named altermagnets \cite{Smejkal2022_2}. Altermagnetism exists in the space between antiferromagnetism and ferromagnetism -- it is characterized by a net zero magnetization, yet the Kramers degeneracy is lifted by a non-relativistic, momentum-dependent spin-splitting of the electronic bands. The band splitting is driven by crystal symmetries connecting opposite-spin sublattices, linked by rotational symmetry in real space \cite{Smejkal2022_1}. Due to intrinsic time-reversal symmetry breaking predicted to emerge in these systems, altermagnets have been theorized to exhibit numerous exotic phenomena, such as large anomalous Hall effect, spin-polarized currents and magneto-optical Kerr effect \cite{Smejkal2020, Mazin2022, Smejkal2022_3}.


Since its initial theoretical formulation, altermagnetism and related properties have now been explored in several materials, such as RuO$_2$ \cite{Gonzalez2021, Feng2022, Bai2022, Karube2022, Bose2022, Tschirner2023, Wang2023, Guo2024, Wang2024, Osumi2025, Kebler2024, Hiraishi2024, Fedchenko2024}, MnTe \cite{Krempasky2024, Lee2024, Osumi2024,Betancourt2023, Kluczyk2024, Gonzalez2024,Hariki2024,Liu2024} and others \cite{Han2024, Reichlova2024, Biniskos2022, Nakatsuji2015, Reimers_2024, Ghimire2018, Jiang2024sdw}. MnTe for example has been found to exhibit Kramers degeneracy lifting \cite{Krempasky2024, Lee2024, Osumi2024}, an anomalous Hall effect \cite{Betancourt2023, Kluczyk2024}, anisotropic magnetoresistance \cite{Gonzalez2024} and time-reversal symmetry breaking \cite{Hariki2024}. As altermagnetism is generally driven by specific crystal symmetries connecting different spin-sublattices, the phenomena discovered thus far do not necessarily require many-body interactions for their emergence \cite{Smejkal2022_2}. Therefore, one of the central thrusts in this budding field remains whether strongly correlated electronic phases prevalent in other families of quantum materials, such as superconductivity, various density waves or electronic nematicity, can be realized and studied in altermagnets, and how they can be controlled \cite{Smejkal2022_2}. This has however remained challenging. Moreover, there has been little atomic-scale insights exploring the family of altermagnets.


Recently, a quasi-2D layered system, Co-intercalated $2H$-NbSe$_2$ (Co$_{0.25}$NbSe$_2$), emerged as another altermagnetic candidate \cite{Regmi2024}. Experiments, supported by theoretical calculations, showed that it exhibits a spin-polarized electronic band structure consistent with altermagnetism \cite{Dale_2024,Regmi2024, deVita_switch}. Given that the ``parent'' material $2H$-NbSe$_2$ hosts the canonical 3$a_0$ $\times$ 3$a_0$ charge density wave and superconductivity \cite{NbSe2_3x3_CDW, Giambattista1988NbSe2, Hess1989NbSe2vortex}, Co intercalation of the superconducting $2H$-NbSe$_2$ is a potentially promising platform in the quest for engineering correlated electron states that coexist with altermagnetism. Indeed, through optical pump-and-probe spectroscopy as well as magnetic susceptibility measurements, two phase transition temperatures have already been identified, one around 150 K and another around 50 K, suggesting a rich electronic environment beyond pure altermagnetism \cite{Regmi2024, deVita_switch}. In this work, we study altermagnet Co$_{0.25}$NbSe$_2$ using a combination of low-temperature spectroscopic imaging STM, spin-polarized STM, angle-resolved photoemission spectroscopy (ARPES) and density functional theory (DFT) to provide a fresh insight into these additional electronic phenomena. We observe non-dispersive 2$a_0$ $\times$ 2$a_0$ charge modulations on the Se termination, distinct in wave length from those in the parent compound $2H$-NbSe$_2$. Through ARPES, we are able to visualize replica bands accompanying the 2$a_0 \times$ 2$a_0$ density wave state. Differential conductance d$I$/d$V$ spectra further reveal a partial gap opening centered at the Fermi level, which is not observed in our DFT calculations of the material in the altermagnetic state, suggesting an origin beyond altermagnetism. By performing spin-polarized STM, we detect a spectroscopic change in the density of states, accompanied by a change in the relative intensities of the charge modulations for different STM tip-spin direction. This measurement suggests an additional spin component associated with the modulations. Interestingly, we find that the out-of-plane magnetic field can be used to alter the electronic density-of-states, as well as the amplitude of density wave modulations, which can gradually increase or decrease depending on the direction and strength of the magnetic field. Taken together, our experiments reveal the emergence of magnetic-field-tunable density waves in the presence of altermagnetism in Co$_{0.25}$NbSe$_2$. The apparent tunability can be attributed to spins that are not fully aligned with the $c$-axis, being tilted by the external magnetic field, which can have profound implications on the altermagnetic properties and should be taken into account to fully understand this system.

\section{Results}
\begin{figure}
    \centering
    \includegraphics[width = \textwidth]{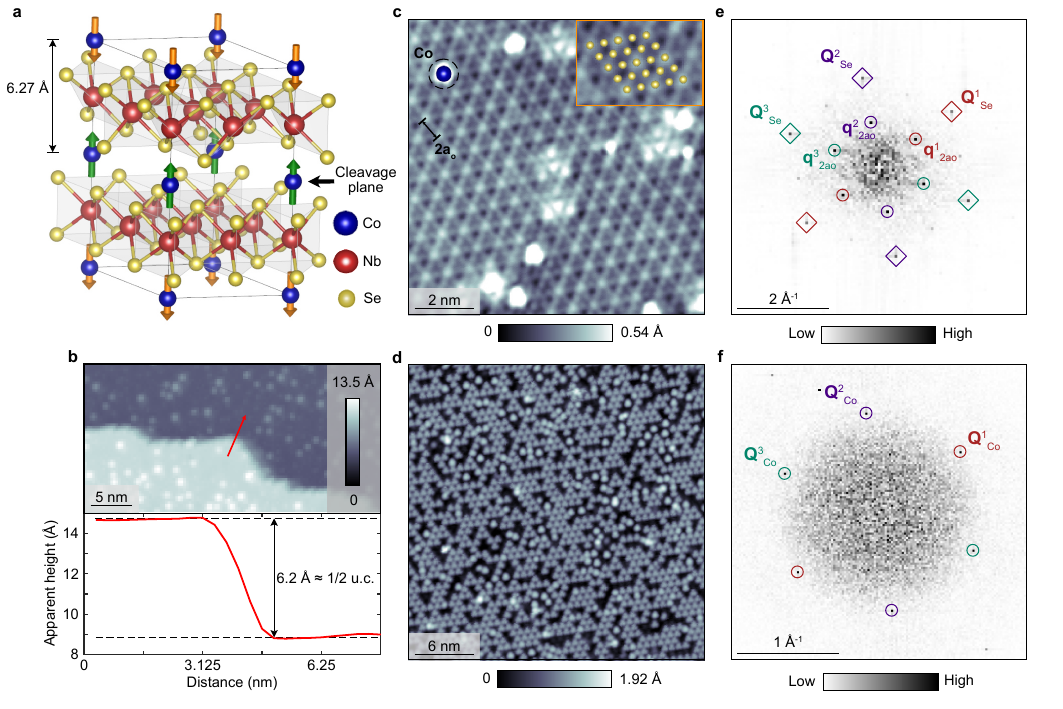}
    \renewcommand{\baselinestretch}{1}
    \caption{\textbf{Crystal structure and surface identification} - \textbf{a}, 3D ball model of the crystal structure of Co$_{0.25}$NbSe$_2$ as viewed from the side. \textbf{b}, (Top) STM topograph of a step edge from a Se-termination to the next Se-termination. (Bottom) A linecut perpendicular to the step-edge along the red line in the top panel, smoothed by the nearest-neighbor averaging. The step height is approximately a half the unit cell shown in (\textbf{a}). \textbf{c, d}, STM topographs of the Se-terminated surface and the Co superstructure, respectively. \textbf{e, f}, Drift-corrected Fourier transforms of the Se surface and Co surface in (c,d), respectively. Circles denote the peaks corresponding to a real-space wave length of $2a_0$, while the diamonds corresponds to the $a_0$ wave length. STM setup conditions: \textbf{b}, $V_{sample}$ = 1 V, $I_{set}$ = 10 pA; \textbf{c}, $V_{sample}$ = 50 mV, $I_{set}$ = 200 pA; \textbf{d},$V_{sample}$ = 100 mV, $I_{set}$ = 10 pA.}
    \label{fig:1}
\end{figure}

We cleave the samples in ultra-high vacuum at cryogenic temperature and immediately insert them into the microscope head (Methods). The crystals have a 2D layered structure, with interstitial Co atoms residing between adjacent Se-Nb-Se slabs, above the Nb atoms (Fig.\ \ref{fig:1}a). The topographic height between adjacent layers is 6.2 $\si{\angstrom}$ (Fig.\ \ref{fig:1}b), which matches the expected interlayer distance extracted from the diffraction measurements (Fig.\ \ref{fig:1}a). Typical STM topographs show a hexagonal lattice with two possible surface terminations: the Se termination in which Se atoms form a hexagonal lattice with the lattice constant $a_0 \approx 3.7 \si{\angstrom}$ (Fig.\ \ref{fig:1}c), or the Co termination where some of the interstitial Co atoms are left on top of the Se surface (Fig.\ \ref{fig:1}d). The remaining Co atoms on the surface exhibit a tendency to order into a 2 $\times$ 2 superstructure, consistent with their expected structure in the bulk between the adjacent NbSe$_2$ slabs \cite{Regmi2024}. Interestingly, STM topographs of the Se termination also show additional modulations that are approximately commensurate with the Se lattice, where every other atom appears brighter than the adjacent ones in all in-plane lattice directions (inset in Fig.\ \ref{fig:1}c). This can be seen in the corresponding Fourier transforms (FTs) of STM topographs where, in addition to the Se atomic Bragg peaks, another set of superstructure peaks with wave vectors $\textbf{q}_{2a0}^i = \frac{1}{2} \textbf{Q}_{Se}^i$ ($i$ = 1, 2, or 3 labels different in-plane lattice direction) appears (Fig.\ \ref{fig:1}e). We have observed these 2 $\times$ 2 modulations on 5 samples with 8 different STM tips. The 3$a_0$ $\times$ 3$a_0$ charge density wave state known to exist in the bulk and on the Se surface of undoped NbSe$_2$ at about $\frac{1}{3}\textbf{Q}^i_{Se}$ \cite{NbSe2_3x3_CDW} is notably absent in Co$_{0.25}$NbSe$_2$ (Fig.\ \ref{fig:1}e,f). The 2$a_0$ $\times$ 2$a_0$ modulations detected on the Se surface are centered directly on top of Se atoms (Fig.\ \ref{fig:1}c). This is laterally offset from the Co atom positions, which are centered under every other Nb atom in the layer below (Fig.\ \ref{fig:1}a). 

\begin{figure}
    \centering
    \includegraphics[width = \textwidth]{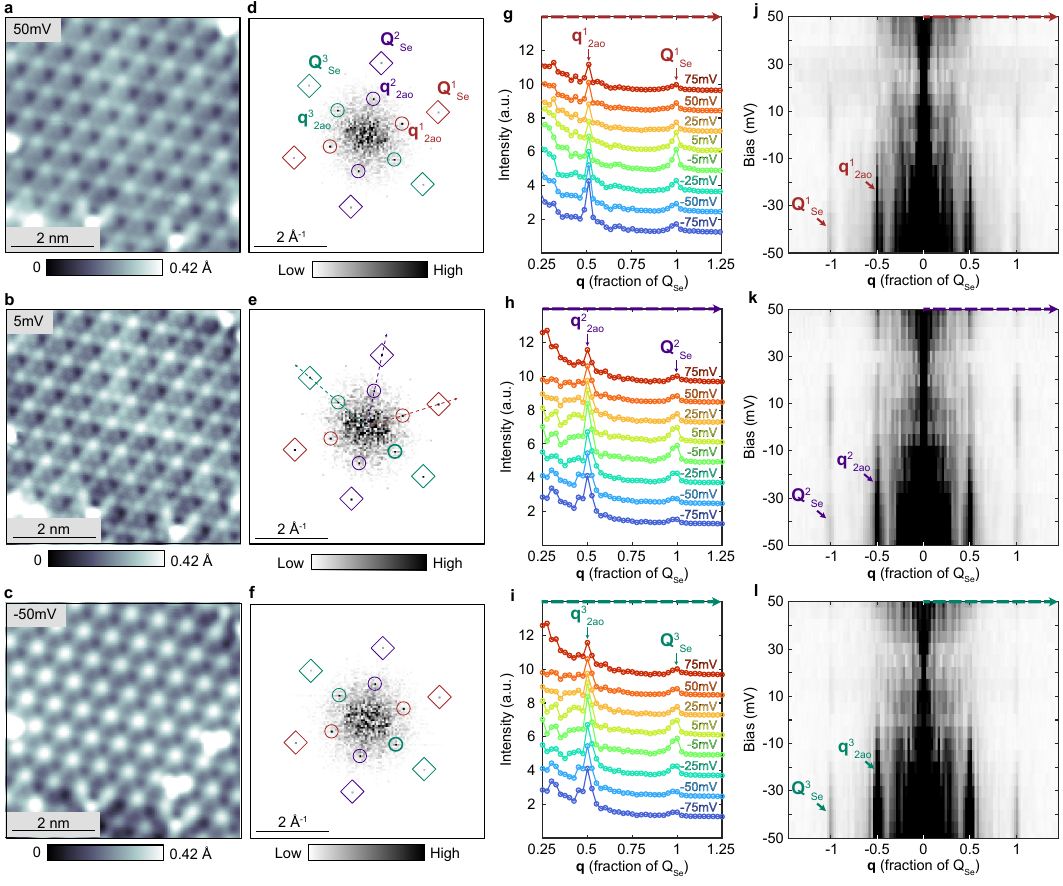}
    \renewcommand{\baselinestretch}{1}
    \caption{\textbf{Energy dependence of charge modulations.} - \textbf{a-c}, Bias dependent STM topographs of the Se surface, and \textbf{d-f} associated Fourier transforms. \textbf{g}, Linecut taken along $\mathbf{Q}^1_{Se}$ as defined by red dotted line in (\textbf{e}) for different biases, with $\mathbf{q}^1_{2ao}$ and $\mathbf{Q}^1_{Se}$  identified. \textbf{h, i}, Same as (\textbf{g}) but taken along $\mathbf{Q}^2_{Se}$ and $\mathbf{Q}^3_{Se}$, respectively. \textbf{j}, Waterfall plot of the FT linecuts, starting from one $\mathbf{Q}^1_{Se}$ to its mirror image, taken from a density of states map. \textbf{k,l}, Same as (\textbf{j}), but taken along $\mathbf{Q}^2_{Se}$ and $\mathbf{Q}^3_{Se}$, respectively. STM setup conditions: \textbf{a}, $V_{sample}$ = 50 mV, $I_{set}$ = 100 pA; \textbf{b}, $V_{sample}$ = 5 mV, $I_{set}$ = 40 pA; \textbf{c}, $V_{sample}$ = -50 mV, $I_{set}$ = 100 pA; From the density of states map used in (\textbf{j-l}), $V_{sample}$ = 50 mV, $I_{set}$ = 100  pA, $V_{exc}$ = 5 mV (rms).}
    \label{fig:2}
\end{figure}

In this work, we primarily focus on characterizing the Se surface termination. To gain further insight into the observed charge modulations, we acquire a series of STM topographs and differential conductance d$I$/d$V$(\textbf{r},$V$) maps in a range of biases (Fig.\ \ref{fig:2}). The same $\textbf{Q}_{2a0}^i = \frac{1}{2} \textbf{Q}_{Se}^i$ peaks can be seen in the FTs of all STM topographs acquired with a bias between -500 and +500 meV (Fig. \ref{fig:2}a-i, Extended Data Figures 1,2). Similarly, $\textbf{Q}^i_{CO}$ can also be detected in d$I$/d$V$(\textbf{r},$V$) maps in a range of different biases (Fig.\ \ref{fig:2}j-l). The wave vectors are extremely localized in reciprocal space ($\sim$1-2 pixels width or 0.007 - 0.014 $\si{\angstrom}^{-1}$ width) and non-dispersive with energy. On this basis, we attribute these features to a static 2$a_0$ $\times$ 2$a_0$ charge ordering along all in-plane lattice directions. We can exclude electron scattering and interference as the origin of observed modulations, as these are typically broader and result in Fourier peaks that disperse, or change their reciprocal space position, as a function of energy \cite{Hoffman2002f, Aynajian2012, Zeljkovic2014, Arguello2015, Gao2018, Fujita2019}.

The 2$a_0$ $\times$ 2$a_0$ charge modulations remain visible in STM data up to at least 50 K, which was the highest temperature accessible in our STM experiment (Extended Data Figure 3). To determine the approximate temperature scale associated with the emergence of the 2 $\times$ 2 modulations, we turn to the ARPES measurements of the Se surface (Fig.\ \ref{fig:3}). We first mention that the constant energy contour at the Fermi level around $\Gamma$ point determined by ARPES presents a good match to those obtained by DFT calculations of the system in the altermagnetic state (Fig.\ \ref{fig:3}a, c-e, Methods), which is discussed in more detail in Ref.\ \cite{deVita_switch}. Since 2$a_0$ $\times$ 2$a_0$ modulations should in principle result in folding of the electronic bands by the same wave vector, we look for signatures of band replicas in our data. At 20 K, it can be seen that the flower-like $\Gamma$ point feature is also seen duplicated at the $M$ point (circled in Fig.\ \ref{fig:3}c). This feature at $M$ is substantially more faint in the equivalent plot at 85 K, and seems to vanish in the 150 K measurement (Fig.\ \ref{fig:3}d,e). The absence of band replicas above this temperature also suggests that the 2 $\times$ 2 structural ordering of Co atoms beneath the topmost NbSe$_2$ layer imaged is unlikely to be driving the observed band folding, as this should in principle remain present across all temperatures. A phase transition around 150 K is also confirmed by pump-and-probe optical spectroscopy with a sign-change in the integrated reflectivity \cite{deVita_switch}.

A notable difference between the integrated density of states spectrum from surface-projected DFT calculations and our large-scale STM d$I$/d$V$ spectrum is that DFT predicts no prominent spectral gap at the Fermi level in the altermagnetic state of the system (Fig.\ \ref{fig:3}f,g). However our d$I$/d$V$ spectra detect a partial V-shaped spectral gap $\Delta \approx$ 30 meV centered at the Fermi level (Fig.\ \ref{fig:3}g). Thus, some qualitative difference must be present between the DFT optimized structure and STM experiments. We find that the gap is generally homogeneous across the sample and impurities (Fig. \ref{fig:3}h,i). The spectra show a substantial amount of residual conductance at zero energy (Fig.\ \ref{fig:3}g,h,i), suggesting that the Fermi surface is not fully gapped at low temperature. 

It is interesting to further examine if there are any spin characteristics associated with the density wave modulations. For this purpose, we utilize spin-polarized STM tips such that spin polarization of the tip can be controlled by external magnetic field (Extended Data Figure 4, Methods). We repeat the equivalent experiment as before, acquiring topographs and d$I$/d$V$(\textbf{r},$V$) maps (Fig.\ \ref{fig:4}a-c) as a function of magnetic field applied parallel or antiparallel to the $c$-axis. Let us first focus on the intensity of the three $|\textbf{Q}^i_{2a0}|$ peaks in STM topographs acquired in magnetic field applied in the opposite directions (Fig.\ \ref{fig:4}e). We can observe systematic differences in the intensities for positive magnetic fields, when the tip spin polarization is ``up'', compared to negative magnetic fields when the tip's spin polarization is ``down'' (Fig.\ \ref{fig:4}e). Examining this behavior in more detail, a similar asymmetry can be seen in FTs of d$I$/d$V$(\textbf{r},$V$) maps as well. Dispersions of the intensities of $|\textbf{Q}^i_{2a0}|$ as a function of bias $V$ appear distinctly different for opposite directions of the magnetic field (Fig.\ \ref{fig:4}f-h). The change is systematic across many fields swept and different bias setup conditions used (Extended Data Figures 5 and 6). Sensitivity of the density wave modulation amplitude to the spin polarization of the tip probe suggests that there exists a spin modulation accompanying the charge modulations (Fig.\ \ref{fig:4}i). Fig.\ \ref{fig:4}j shows a simple schematic of coexisting spin and charge density waves on a triangular lattice consistent with our experimental data.

Lastly, we turn to another intriguing aspect of our spin-polarized STM data. As we examine the evolution of d$I$/d$V$ spectra more closely, we observe a continuous change in the intensity at higher magnetic fields (Fig.~\ref{fig5}a). $|\textbf{Q}^i_{2a0}|$ Fourier transform amplitudes also exhibit fine magnetic-field-induced changes at higher fields (Fig.~\ref{fig5}c-e). These cannot be explained by a gradual change of the STM tip polarization direction only, as the tip's spin-direction should already be fully aligned with the external field by a few Tesla. As such, it appears that out-of-plane magnetic field modifies the electronic density-of-states of the sample as well as the density modulations. To better understand if these magnetic-field-induced changes concomitantly alter electronic properties in the pure charge channel, we repeat equivalent measurements with a non-spin-polarized STM tip. Interestingly, we find a smaller, but nevertheless sizable, response in the electronic density-of-states, resulting in a spectral change of opposite magnitude when magnetic field reverses direction (Fig.~\ref{fig5}b). The amplitudes of $|\textbf{Q}^i_{2a0}|$ show minimal changes as a function of the fields taken (Fig.~\ref{fig5}f-h). Overall, these measurements suggest a magnetic-field tunable nature of the electronic density-of-states, with a particularly pronounced effect on the spin density modulations.

\section{Discussion}

While the field of altermagnetism has progressed rapidly, atomic-scale imaging of material candidates has been extremely rare. Our experiments provide an atomic-scale spectroscopic glimpse to reveal 2$a_0$ $\times$ 2$a_0$ modulations with both charge and spin components in Co-intercalated NbSe$_2$. Such observations could provide explanations for the multiple temperature transitions seen by other experiments \cite{deVita_switch, Regmi2024}. Given the emergent theoretical framework of unconventional Cooper pair density waves and SDWs in altermagnets \cite{Parthenios2025DWtheory, Sim2024PDW}, a recent experiment on a $d$-wave altermagnet that reveals signatures of an SDW from macroscopic non-local probes \cite{Jiang2024sdw}, and our nanoscale discovery of charge and spin density waves on a $g$-wave altermagnet, density waves may soon become a widespread phenomenon in altermagnets. Importantly, in our work, we can unambiguously disentangle charge and spin density modulations, facilitated by the atomically-resolved and spin-resolved sensitivity of our measurements. Our experiments further demonstrate substantial tunability by external magnetic field.


Given the asymmetric response of the density waves dependent on the tip's spin polarization, the density modulations detected in our experiments also break the time-reversal symmetry. This is further supported by the spin-polarized density of states measurements that show distinct differences for opposite tip spin polarizations (Fig.\ \ref{fig:4}d). In future work, it would be of particular interest to explore the existence and potential control of time-reversal symmetry breaking domains via spin-polarized STM, complementary to the X-ray photoemission electron microscopy domain studies of altermagnetic MnTe thin films \cite{Amin2024}.  

\begin{figure}
    \centering
    \includegraphics[width = \textwidth]{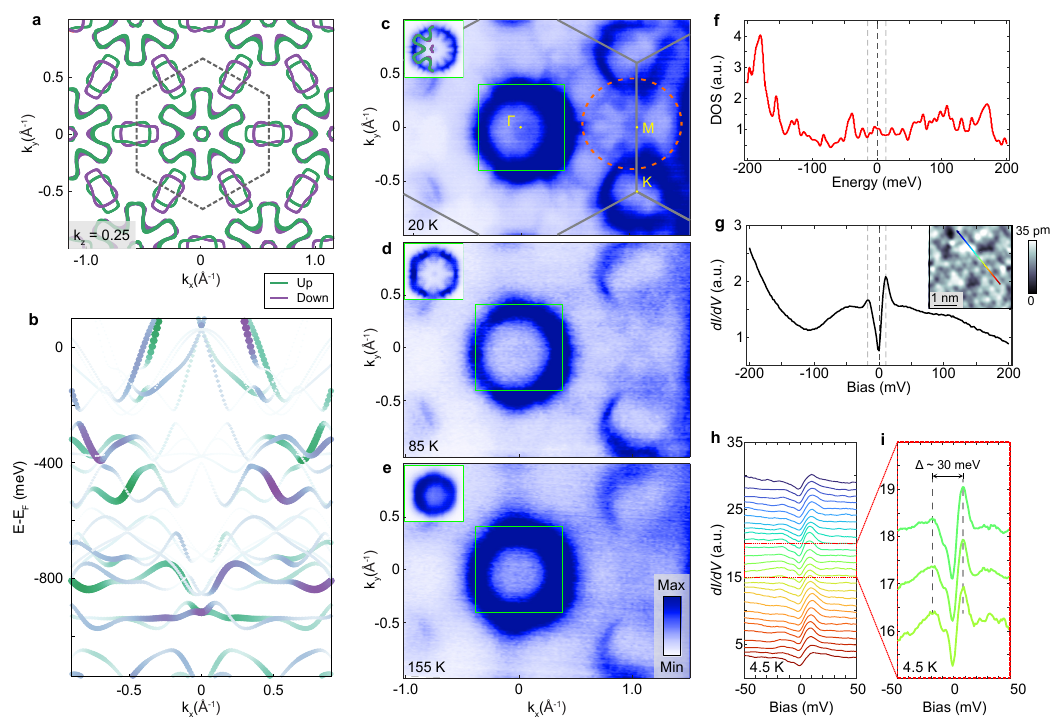}
    \renewcommand{\baselinestretch}{1}
    \caption{\textbf{Theoretical calculations and experimental measurement of electronic density-of-states.} \textbf{a}, Bulk Fermi surface at $k_z$ = 0.25, an intermediate layer in k-space between $k_z$ = 0 and $k_z$ = 0.5, from DFT calculations (Methods). The Brillouin zone of the 2 $\times$ 2 superstructure is denoted by a dashed hexagon. \textbf{b}, Unfolded electronic band structure along the $M'$ - $\Gamma$ - $M'$ direction from DFT calculations. Green and purple colors in (a,b) denote different spins. \textbf{c-e}, Experimental Fermi surfaces acquired using ARPES with $hv$ = 70 eV photon energy at 20 K, 85 K, and 155 K, respectively. Fermi surfaces taken using photon energy $hv$ = 25 eV are included in the insets. The gray hexagon in (c) denotes the original Se-Se Brillouin zone, and the orange circle highlights the band replica. \textbf{f}, DFT-simulation of the Se-projected density of states summed over the Se surface atoms from a simulation of the Se-terminated surface. \textbf{g}, Average d$I$/d$V$ spectrum from STM measurements taken over the Se-terminated region in the inset. \textbf{h}, d$I$/d$V$ spectra taken along the line over the topography in the inset of (\textbf{g}) with next nearest-neighbor averaging. \textbf{i}, A zoom-in on the spectra, showing an energy gap of $\sim$ 30 meV, which appears to be homogeneous. STM setup conditions: \textbf{g}, $V_{sample}$ = -200 mV, $I_{set}$ = 200 pA, $V_{exc}$ = 2 mV (rms); \textbf{g inset}, $V_{sample}$ = 50 mV, $I_{set}$ = 150 pA; \textbf{h}, $V_{sample}$ = 50 mV, $I_{set}$ = 150 pA, $V_{exc}$ = 1 mV (rms).}
    \label{fig:3}
\end{figure}

\begin{figure}
    \renewcommand{\thefigure}{4}  
    \renewcommand{\figurename}{FIG.}  
    \centering
    \includegraphics[width = \textwidth]{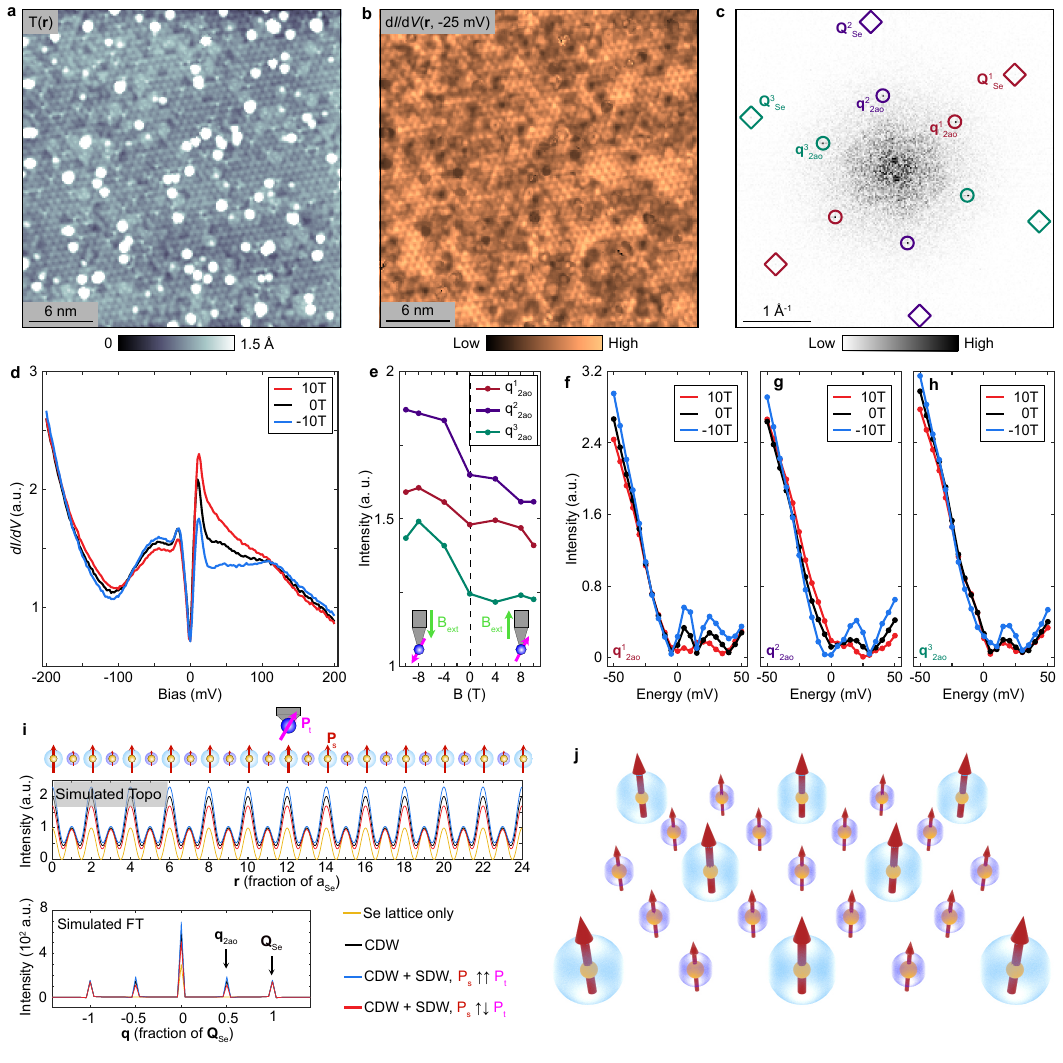}
    \renewcommand{\baselinestretch}{1}
    \caption{\textbf{Characterization of spin modulations using spin-polarized STM.} -  \textbf{a}, STM topograph at 0 T, \textbf{b}, $dI/dV$ map of the same area, and \textbf{c}, its Fourier transform. Density wave peaks and the atomic Bragg peaks are defined by the same scheme as that in Fig.\ 1. \textbf{d}, d$I$/d$V$ spectra of the Se surface at 0 T (black), 10 T (red), and -10 T (blue), taken over the same area as that in Fig.\ 3h. Minus sign denotes the reversal of magnetic field along the $c$-axis. \textbf{e}, Intensities of the three charge modulation peaks from the Fourier transform of STM topographs taken at different fields with a spin-polarized tip. Inset cartoon schematics show spin-polarized tip alignment with external magnetic field. \textbf{f-h}, Intensities of the charge modulation peaks $\mathbf{q}_{2a0}^1$, $\mathbf{q}_{2a0}^2$, and $\mathbf{q}_{2a0}^3$ in d$I$/d$V$ maps as a function of energy for different magnetic fields. \textbf{i}, A pedagogical 1D simulation of the charge and spin modulations. The cartoon at the top shows the Se atoms (yellow circles) with their surrounding electron cloud (blue circles) alternating in size denoting the charge density wave (CDW) modulation, and magnetic moments $P_s$ (red arrows) alternating in size, showing the}
\label{fig:4}
\end{figure}

We stress that charge modulations observed here have a strong electronic component, as opposed to a purely structural, for the following reasons: (1) charge modulations are centered on the Se sites, which is away (in the $ab$-plane) from the 2 $\times$ 2 ordered Co atoms located under the topmost NbSe$_2$ slab; (2) spin-polarized STM measurements suggest there is also an underlying spin modulation associated with the order; (3) band replicas in ARPES measurements are only visible below an intermediate temperature of $\sim$ 100-150 K (Fig.\ \ref{fig:3}c-e), ruling out the structural arrangement of Co atoms as low energy electron diffraction shows 2 $\times$ 2 Co super-lattice is visible up to much higher temperature; (4) magnetic field tunability by an out-of-plane magnetic field suggests an unusual electronic state beyond simple structural ordering. 

Our experimental observations are consistent with spin modulations that are characterized by spins with some in-plane component, which cant to align with the out-of-plane magnetic field; this explains a strong change in the density modulation amplitude when using spin-polarized tips (Fig.~\ref{fig5}c-e) and a much smaller response in the pure charge channel using non-spin-polarized tips (Fig.~\ref{fig5}f-h). In this scenario, as the spins cant to align with external magnetic field, this can in principle break the spatial symmetry mapping between spin-up and spin-down states, which should in turn affect altermagnetic band splitting. It is conceivable that this is already detected in our experiments, as the electronic band structure measured by d$I$/d$V$ spectra also changes (Fig.~\ref{fig5}a,b). Since spin canting away from the $c$-axis inevitably selects an in-plane direction, this should also lead to C$_6$ rotational symmetry breaking and possible domain formation akin to MnTe \cite{Amin2024}, which would be of interest to explore in subsequent experiments. Further investigations into these canted spins would not only be important for fundamental understanding of the altermagnetic spin textures, but could also be desirable for spintronic applications \cite{Liu2024_2, Leenders2024, Das2022, Lebun2018}. 

While superconductivity is not detected in our samples, the parent system $2H$-NbSe$_2$ without Co intercalation is a well-known bulk superconductor with $\sim$ 3$a_0$ $\times$ 3$a_0$ charge density wave state. Therefore, it would be of particular interest to explore how undoped $2H$-NbSe$_2$ evolves into a non-superconducting altermagnet with a 2$a_0$ $\times$ 2$a_0$ charge and spin modulations at $x=1/4$ Co concentration by studying a range of intermediate Co concentrations. It is conceivable that intercalation of a smaller density of Co atoms, or intercalating a different magnetic element altogether, could stabilize both superconductivity and altermagnetism in the same compound, providing a unique platform to study the interplay of superconductivity and altermagnetism, potentially generating unconventional Cooper pair density waves \cite{Parthenios2025DWtheory}. In general, intercalating transition metal dichalcogenides can provide a widely tunable platform to realize new materials systems, in the search for novel correlated electronic states in altermagnets. \\
\\
\noindent{\bf Author contributions}\\
C.C. performed STM measurements with the help from M.X., S.C. and A.L. F.M., A.D.V., C.B., and J.A.M. collected the ARPES data. M.C. and D.R. performed the DFT calculations. Y.H. synthesized the bulk single crystals. Z.W. provided a theoretical input on the experimental analysis. I.Z., C.C. and F.M. wrote the paper with the input from all the authors. I.Z. supervised the project. \\
\\
\noindent{\bf Methods}\\
\indent \textit{\textbf{Sample growth:}} 

Single crystals of Co$_{0.25}$NbSe$_2$ were grown via the chemical vapor transport (CVT). High-purity powders of cobalt (Co, 99.99$\%$), niobium (Nb, 99.999$\%$), and selenium (Se, 99.9999$\%$) were used as starting materials. To minimize contamination and residual oxygen, the quartz ampoule underwent thorough chemical cleaning and vacuum heat treatment before loading the reactants. The precursor materials were then sealed inside a quartz ampoule (approximately 10 mm in diameter and 150 mm in length) along with iodine (5 mg/cm$^3$ relative to ampoule volume) as a transport agent. After being evacuated to high vacuum, the sealed ampoule was placed in a two-zone horizontal furnace, with the source region maintained at a higher temperature than the deposition zone. Achieving high-quality crystal growth required precise control of the temperature gradient and iodine concentration. For Co${0.25}$NbSe$_2$, the source region was maintained at 960–980 $^\circ$C, while the growth region temperature was systematically increased from 880 $^\circ$C to 900 $^\circ$C over 100 hours. The system was then held at these temperatures for an additional 300 hours to enable the formation of large single crystals. Finally, a controlled cooling process was implemented over 100 hours, lowering the source region to 200 $^\circ$C and the growth region to 100 $^\circ$C before allowing the ampoule to reach room temperature naturally. The obtained crystals had typical dimensions of approximately 5 $\times$ 5 $\times$ 0.1 mm$^3$. Residual iodine was removed by rinsing the crystals with a methanol solution. The composition was preliminarily assessed via energy-dispersive x-ray spectroscopy (EDS) using a field-emission scanning electron microscope (FE-SEM, JEOL 7500).

\textit{\textbf{STM experiments: }}Samples were glued to the sample holder using EPO-TEK H20E silver conducting epoxy and cured at 175 \textdegree C for 20 min. The cleaving rod was then glued to the top of the sample in the same way. We cold-cleaved the crystals in UHV at a cryogenic temperature (approximately few tens of Kelvin) and quickly inserted them into the STM head for scanning. STM data was acquired using a customized Unisoku USM1300 microscope. The STM tips used were homemade, chemically-etched tungsten tips, annealed in UHV to bright orange color prior to STM experiments. For preparing a spin-polarized tip, the tip scanned and bias-pulsed over the sample, which gave rise to the tip picking up Co adatoms and becoming spin-polarized. After acquiring data with the spin polarized tip, its spin polarization was characterized by scanning on a UHV-cleaved surface of FeTe (see Supplementary Figure 4), an antiferromagnet with a well-defined bicollinear AFM ordering on the surface. By this, we found that typical spin-polarized tips are ferromagnetic-like (i.e. change their polarization direction with different directions of magnetic field). Unless otherwise specified, STM measurements were taken at about 4.8 K.

\textit{\textbf{STM analysis:}} To read out the intensity of the 2 $\times$ 2 peaks in the Fourier transform, the Lawler-Fujita drift-correction algorithm was applied to our topographs and DOS maps. \cite{Lawler2010} This algorithm shifts and crops the original image so that the atomic Bragg peaks are single pixel and even integer coordinates in Fourier space, thus minimizing artificial effects from piezo and thermal drift. Since the 2 $\times$ 2 peaks are a multiple of the lattice constant, these too are shifted to single pixels for intensity read-out. 

\textit{\textbf{ARPES experiments}:}
The ARPES data were obtained at the CASSIOPÉE beamline of Synchrotron SOLEIL (France) using linearly horizontal polarized light with 70 eV and 25 eV photon energies. The Fermi surfaces were integrated $\pm 15$ meV from the Fermi level. The samples were cleaved in UHV  at a pressure better than $3 \times 10^{-10}$ mbar and the spectra were collected with a Scienta R4000 analyzer with momentum and energy resolution better than 0.018 $\si{\angstrom}^{-1}$ and 10 meV, respectively.

\textit{\textbf{DFT calculations:}} 
Density Functional Theory (DFT) calculations have been performed in the collinear spin-polarized configuration and via the plane-wave pseudopotential method, as implemented in the Quantum ESPRESSO package~\cite{QE-2009,QE-2017}. Electron-ion interaction has been modeled for Se atoms via a norm-conserving pseudopotential, while for Nb and Co atoms via ultrasoft pseudopotentials: we chose an energy cut-off of $45$ Ry and $450$ Ry for the wave-function and electron density respectively. 

Electronic band structure for the bulk Co doped $2\times2$ NbSe$_2$ supercell in the altermagnetic phase has been obtained through Perdew-Burke-Ernzerhof (PBE)~\cite{PhysRevLett.77.3865} exchange-correlation functional, sampling the Brillouin zone with a $k$-vector mesh of $9\times9\times16$ points and a first-order Methfessel-Paxton~\cite{PhysRevB.40.3616} electronic smearing of $5$ mRy. Electronic dispersion has then been mapped into the large Brillouin zone of undoped $1\times1$ NbSe$_2$ via an unfolding procedure~\cite{PhysRevB.85.085201} implemented in the code unfold.x~\cite{10.1063/5.0047266}.

The electronic density of states (DOS) projected on the Se atoms was then calculated for the surface, employing a slab geometry consisting of 5 layers of $2H$-NbSe$_2$ and four Co atoms. In order to properly simulate a 2D system we added 20 $\AA$ of vacuum in the non-periodic direction orthogonal to the surface. Ground-state electronic density was then obtained via a $k$-vector mesh of $9\times9\times1$ points and a first-order Methfessel-Paxton~\cite{PhysRevB.40.3616} electronic smearing of $5$ mRy, while DOS has been computed with a $k$-vector mesh of $54\times54\times1$ points and a Gaussian electronic smearing of $0.25$ mRy. 
\\

\noindent{\bf Acknowledgments}\\
I.Z. gratefully acknowledges the support from the US Department of Energy grant number DE-SC0025005. F.M. greatly acknowledges the NFFA-DI funded by the European Union – NextGenerationEU, M4C2, within the PNRR project NFFA-DI, CUP B53C22004310006, IR0000015. M.B.P. and J.A.M gratefully acknowledge support from DanScatt (7129-00018B). D.R. acknowledges support from the HPC resources of IDRIS, CINES, and TGCC under Allocation No. 2024-A0160914101 made by GENCI. This work was supported by the National Research Foundation of Korea (NRF) funded by the Ministry of Education, Science and Technology (NRF-2019M2C8A1057099 and NRF-2022R1I1A1A01063507).\\
\\
\noindent{\bf Data availability}\\
The data that support the findings of this study are available from the corresponding authors upon reasonable request.\\
\\
\noindent{\bf Code availability}\\
The code that supports the findings of the study is available from the corresponding authors upon reasonable request.\\
\\
\noindent{\bf Correspondence and requests for materials} should be addressed to
\textit{younghh@uc.ac.kr}, \textit{federico.mazzola@spin.cnr.it} and \textit{ilija.zeljkovic@bc.edu}.\\
\\
\noindent{\bf Competing financial interests}\\
The authors declare no competing financial interests.\\

\bibliographystyle{custom-style.bst}
\bibliography{alter.bib}


\newpage

\begin{figure}
    \renewcommand{\thefigure}{}
    \renewcommand{\figurename}{(continued)}
    \centering
    \renewcommand{\baselinestretch}{1}
    \caption{spin density modulation. The tip is spin polarized along $P_t$ direction (pink arrow). The top plot shows a simulated linecut along the atomic chain for 4 different types of topographs: Se lattice only (yellow), CDW only (black), combined CDW/SDW where $P_s$ is parallel to $P_t$ (blue), and combined CDW/SDW where $P_s$ is antiparallel to $P_t$. Their corresponding Fourier transforms are represented in the bottom plot following the same color scheme. One can see how the combined CDW/SDW state changes the intensity of the 2$a_0$ modulations and associated Fourier peak $\textbf{q}_{2a0}$. \textbf{i}, A 3D rendering of the coexisting CDW-SDW, with the yellow spheres showing the Se atoms, the light and dark blue spheres showing the electron clouds with a magnetic moment defined by red arrows. STM setup conditions: \textbf{a, b, e-h}, $V_{sample}$ = 50 mV, $I_{set}$ = 150 pA, $V_{exc}$ = 5 mV (rms); \textbf{d}, $V_{sample}$ = -200 mV, $I_{set}$ = 200 pA, $V_{exc}$ = 2 mV (rms).
    \\
    \\
    \\
    \\
    \\
    \\
    \\}
\end{figure}

\begin{figure}
    \renewcommand{\thefigure}{5}
    \renewcommand{\figurename}{FIG.}
    \centering
    \includegraphics[width = \textwidth]{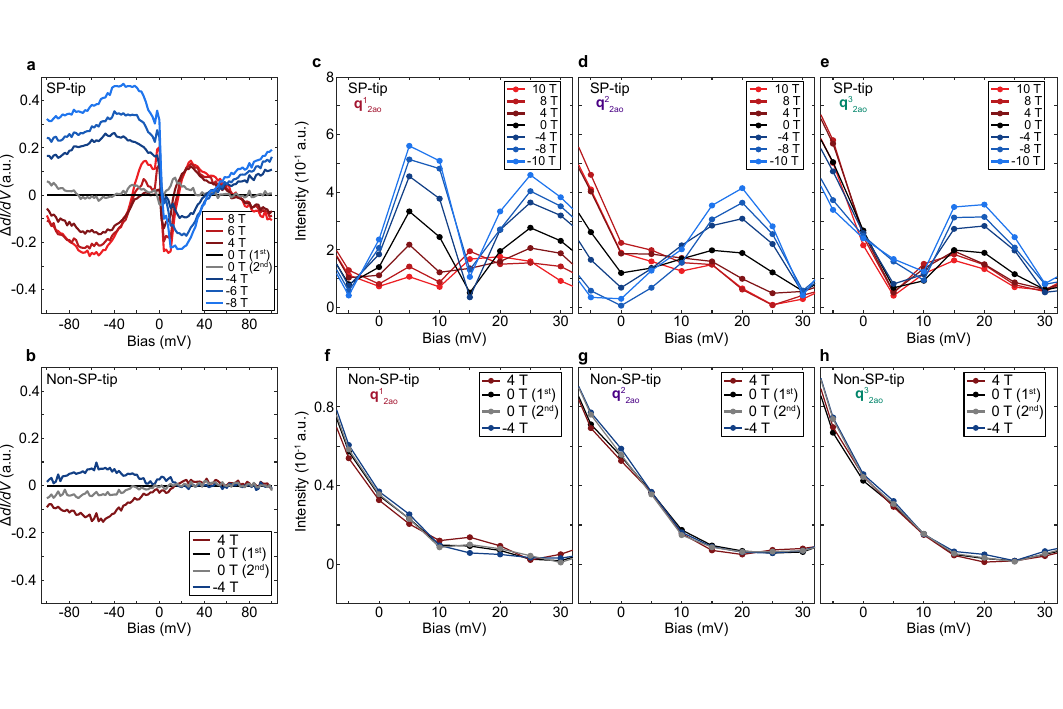}
    \renewcommand{\baselinestretch}{1}
    \caption{\textbf{Magnetic tunability of the electronic density-of-states and the density wave modulations.} - \textbf{a}, Difference in spectra $(\Delta (dI/dV))$ defined as the spectrum at some field subtracted from the first 0 T measurement taken $(dI/dV (B_z) - dI/dV(0 T\, (1^{st}))$. Spectra were acquired with a spin-polarized tip. \textbf{b}, Same as (\textbf{a}), but with data acquired with a non-spin-polarized tip. \textbf{c-e}, Intensity of the 2$a_0$ peaks in Fourier space as a function of sample bias for different magnetic fields along $\mathbf{q}_{2a0}^1$, $\mathbf{q}_{2a0}^2$, and $\mathbf{q}_{2a0}^3$, respectively acquired with a spin-polarized tip. \textbf{f-h}, The same as (\textbf{c-e}), but with data acquired with a non-spin-polarized tip. STM setup conditions: \textbf{a}, $V_{sample}$ = 100 mV, $I_{set}$ = 300 pA, $V_{exc}$ = 2 mV (rms); \textbf{b}, $V_{sample}$ = 100 mV, $I_{set}$ = 300 pA, $V_{exc}$ = 2 mV (rms); \textbf{c-e}, $V_{sample}$ = 50 mV, $I_{set}$ = 150 pA, $V_{exc}$ = 5 mV (rms); \textbf{f-h}, $V_{sample}$ = 50 mV, $I_{set}$ = 150 pA, $V_{exc}$ = 5 mV (rms)}
    \label{fig5}
\end{figure}

\begin{figure}
    \renewcommand{\thefigure}{1}  
    \renewcommand{\figurename}{Extended Data FIG.}
    \centering
    \renewcommand{\baselinestretch}{1}
    \includegraphics[width = \textwidth]{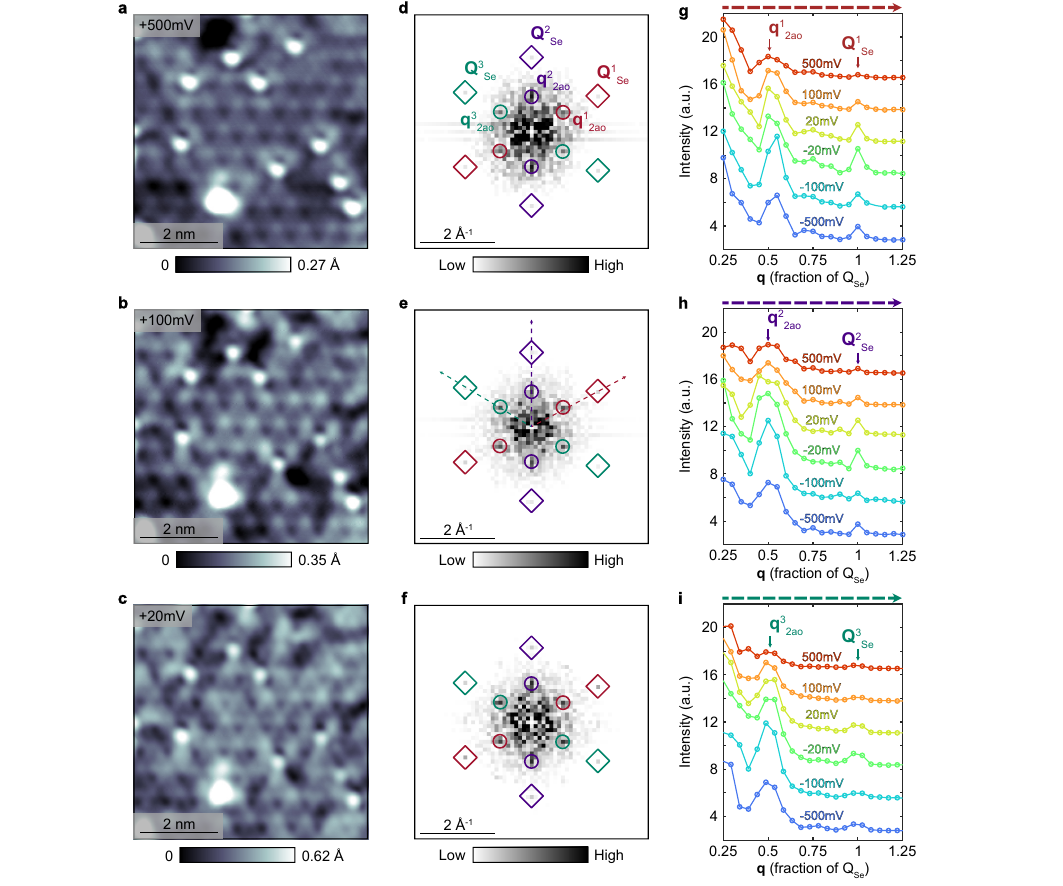}
    \caption{\textbf{Additional energy dependence data of charge modulations.} - \textbf{a-c}, Bias dependent topographs of the Se surface at $V_{sample}$ = 500 mV, 100 mV, and 20 mV, respectively, and \textbf{d-f} associated Fourier transforms. \textbf{g}, Linecut taken along $\mathbf{Q}^1_{Se}$ as defined by the red dotted line in (\textbf{e}) for different biases, with $\mathbf{q}^1_{2a0}$ and $\mathbf{Q}^1_{Se}$  identified. \textbf{h, i}, Same as (\textbf{g}) but taken along $\mathbf{Q}^2_{Se}$ and $\mathbf{Q}^3_{Se}$, respectively. STM setup conditions: \textbf{a}, $V_{sample}$ = 500 mV, $I_{set}$ = 600 pA; \textbf{b}, $V_{sample}$ = 100 mV, $I_{set}$ = 200 pA; \textbf{c}, $V_{sample}$ = 20 mV, $I_{set}$ = 120 pA.}
    \label{suppfig:1}
\end{figure}

\begin{figure}
    \renewcommand{\thefigure}{2}  
    \renewcommand{\figurename}{Extended Data FIG.}
    \centering
    \renewcommand{\baselinestretch}{1}
    \includegraphics[width = \textwidth]{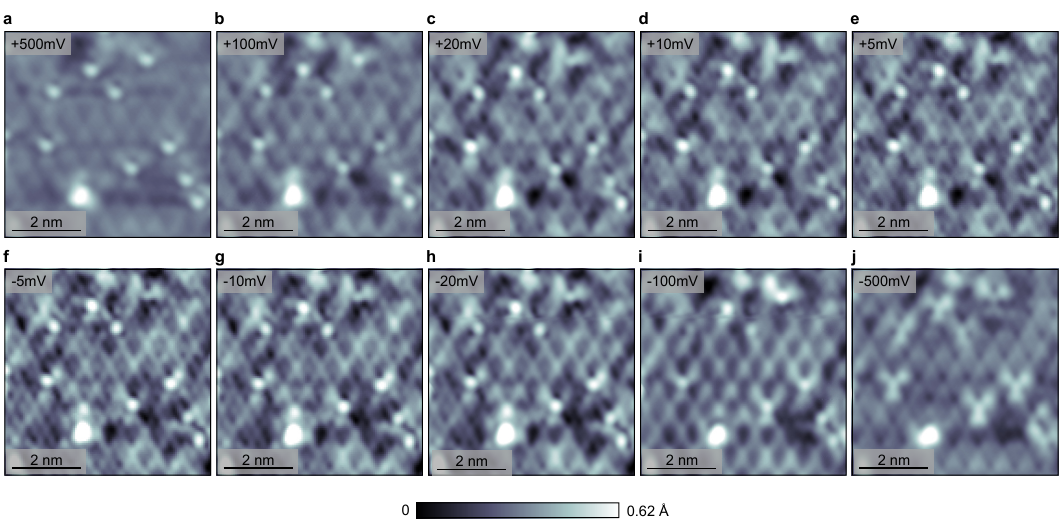}
    \caption{\textbf{Complete data set of bias dependent STM topographs of the Se surface.} STM setup conditions: \textbf{a}, $V_{sample}$ = 500 mV, $I_{set}$ = 600 pA; \textbf{b}, $V_{sample}$ = 100 mV, $I_{set}$ = 200 pA; \textbf{c}, $V_{sample}$ = 20 mV, $I_{set}$ = 120 pA; \textbf{d}, $V_{sample}$ = 10 mV, $I_{set}$ = 110 pA; \textbf{e}, $V_{sample}$ = 5 mV, $I_{set}$ = 105 pA; \textbf{f}, $V_{sample}$ = -5 mV, $I_{set}$ = 105 pA; \textbf{g}, $V_{sample}$ = -10 mV, $I_{set}$ = 110 pA; \textbf{h}, $V_{sample}$ = -20 mV, $I_{set}$ = 120 pA; \textbf{i}, $V_{sample}$ = -100 mV, $I_{set}$ = 200 pA; \textbf{j}, $V_{sample}$ = -500 mV, $I_{set}$ = 600 pA.}
    \label{suppfig:2}
\end{figure}

\begin{figure}
    \renewcommand{\thefigure}{3}  
    \renewcommand{\figurename}{Extended Data FIG.}
    \centering
    \renewcommand{\baselinestretch}{1}
    \includegraphics[width = \textwidth]{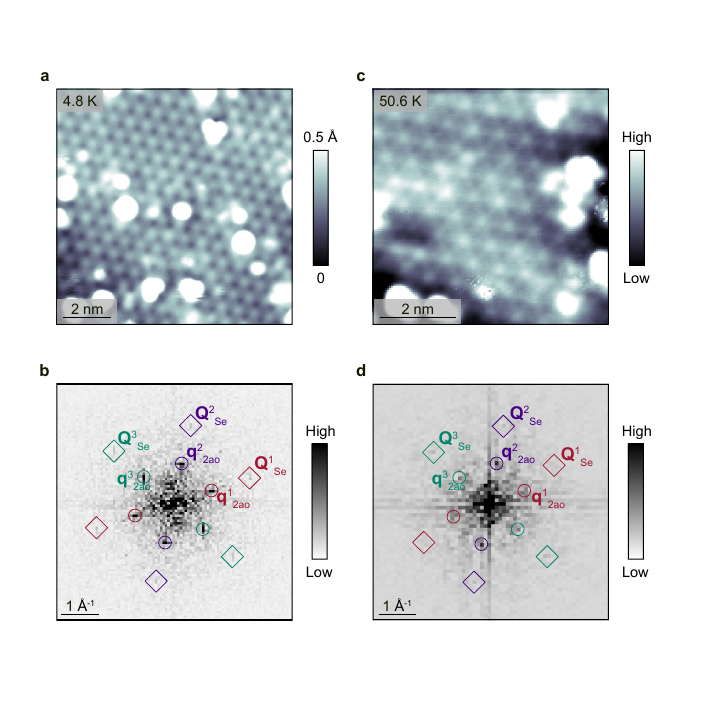}
    \caption{\textbf{STM topographs as a function of temperature} - \textbf{a}, STM topograph of the Se surface at 4.8 K with \textbf{b}, corresponding Fourier transform. \textbf{c}, STM topograph of the Se surface at 50.6 K with \textbf{d}, corresponding Fourier transform. From the Fourier transform, both the Se atomic peaks and 2$a_0$ charge modulation peaks are still present above 50 K. STM setup conditions: \textbf{a}, $V_{sample}$ = 50 mV, $I_{set}$ = 10 pA, \textbf{c}, $V_{sample}$ = 150 mV, $I_{set}$ = 50 pA.}
    \label{suppfig:3}
\end{figure}

\begin{figure}
    \renewcommand{\thefigure}{4}  
    \renewcommand{\figurename}{Extended Data FIG.}
    \centering
    \renewcommand{\baselinestretch}{1}
    \includegraphics[width = \textwidth]{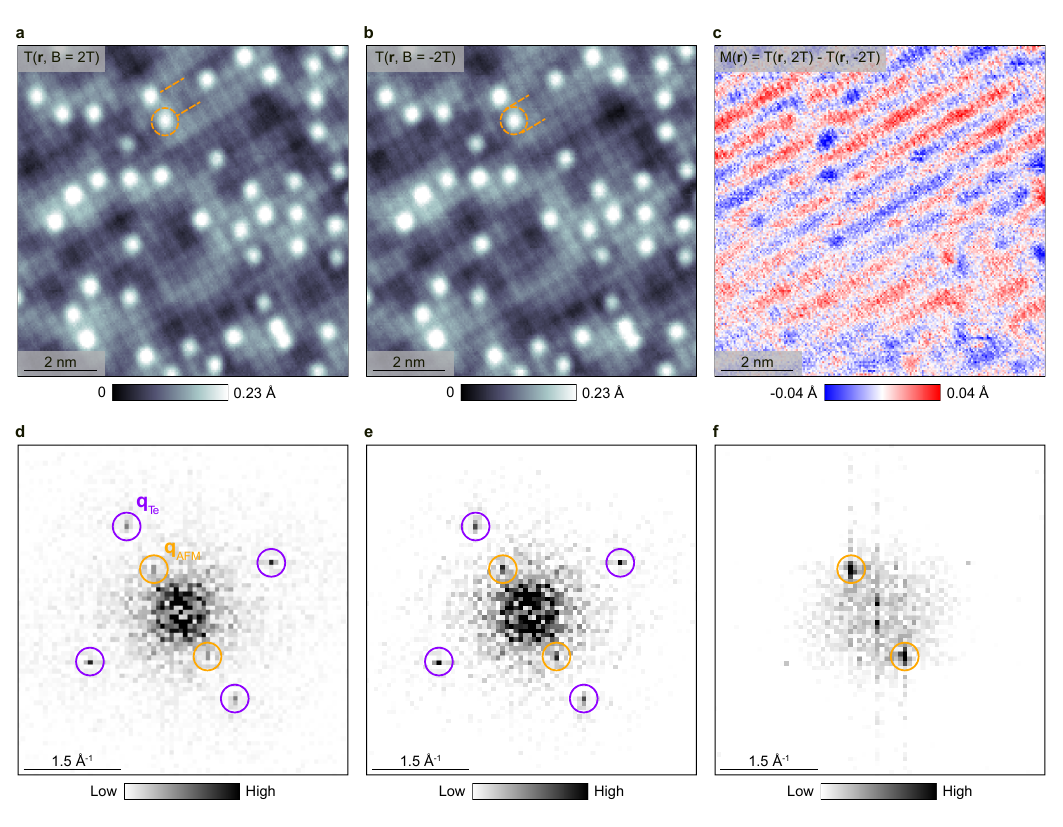}
    \caption{\textbf{Post-characterization of tip's spin polarization.} - \textbf{a,b} STM topographs of the Te termination of cleaved FeTe at (a) 2 T, and (b) -2 T applied perpendicular to the surface (minus sign denotes the reversal of magnetic field direction). The 2$a_0$ stripe order is highlighted with the orange dotted line, which has shifted by $\pi$ due to the tip's spin polarization being reversed by magnetic field. FeTe's antiferromagnetic ordering is not substantially affected by these low fields. \textbf{c}, A spin-resolved magnetic contrast $M$(\textbf{r}) map obtained by subtracting the STM topographs in (a,b). This further highlights the bicollinear antiferromagnetic order, while subtracting out the structural information (the Te lattice). This is due to the spin-polarized tip having a ferromagnetic-like behavior. \textbf{d-f}, corresponding Fourier transforms of (\textbf{a,b,c}), respectively. The Te atomic peaks are circled in purple while the 2$a_0$ order is circled in orange. STM setup conditions: \textbf{a-b}, $V_{sample}$ = 1 V, $I_{set}$ = 10 pA.}
    \label{suppfig:4}
\end{figure}

\begin{figure}
    \renewcommand{\thefigure}{5}  
    \renewcommand{\figurename}{Extended Data FIG.}
    \centering
    \renewcommand{\baselinestretch}{1}
    \includegraphics[width = \textwidth]{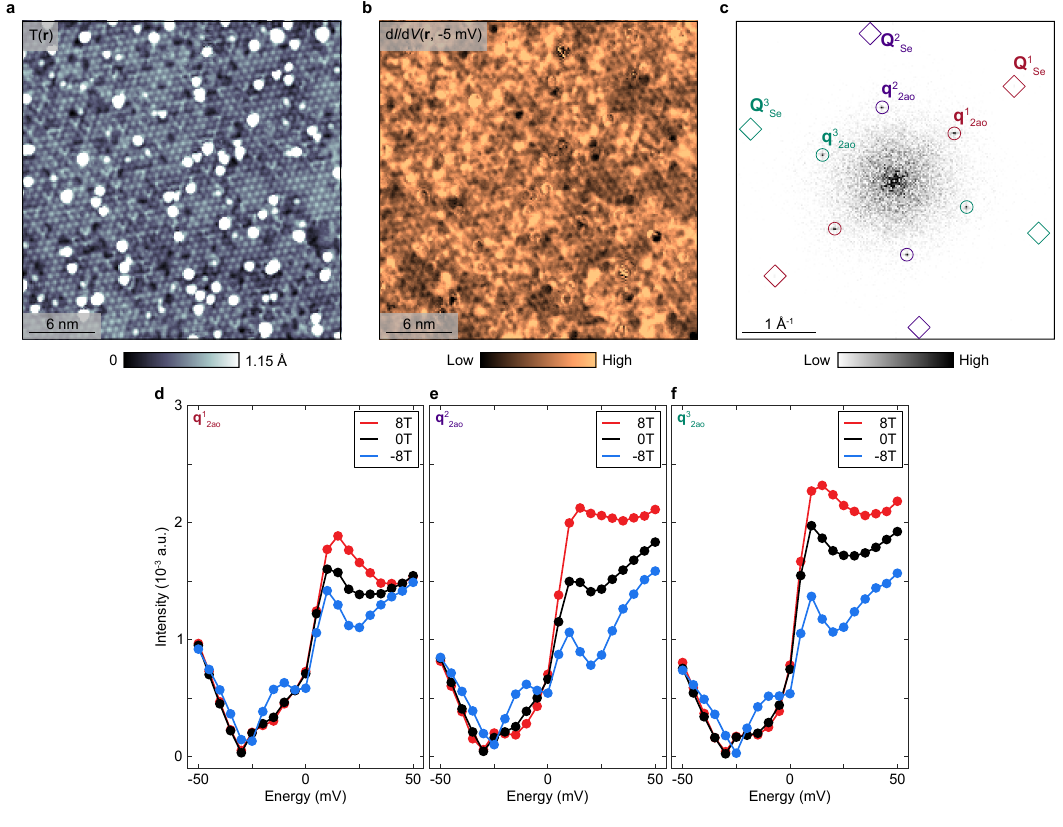}
    \caption{\textbf{Characterization of spin modulations at different settings.} -  \textbf{a}, STM topograph at 8 T, \textbf{b}, $dI/dV$ map of the same area, and \textbf{c}, its Fourier transform over the same area as in Fig. 4, but with $V_{sample}$ = -50 mV instead of +50 mV.  \textbf{d-f}, Intensities of the charge modulation peaks $\mathbf{q}_{2a0}^1$, $\mathbf{q}_{2a0}^2$, and $\mathbf{q}_{2a0}^3$ in d$I$/d$V$ maps as a function of energy for different magnetic fields taken with a spin-polarized tip. STM setup conditions: $V_{sample}$ = -50 mV, $I_{set}$ = 150 pA.} 
    \label{suppfig:5}
\end{figure}

\begin{figure}
    \renewcommand{\thefigure}{6}  
    \renewcommand{\figurename}{Extended Data FIG.}
    \centering
    \renewcommand{\baselinestretch}{1}
    \includegraphics[width = \textwidth]{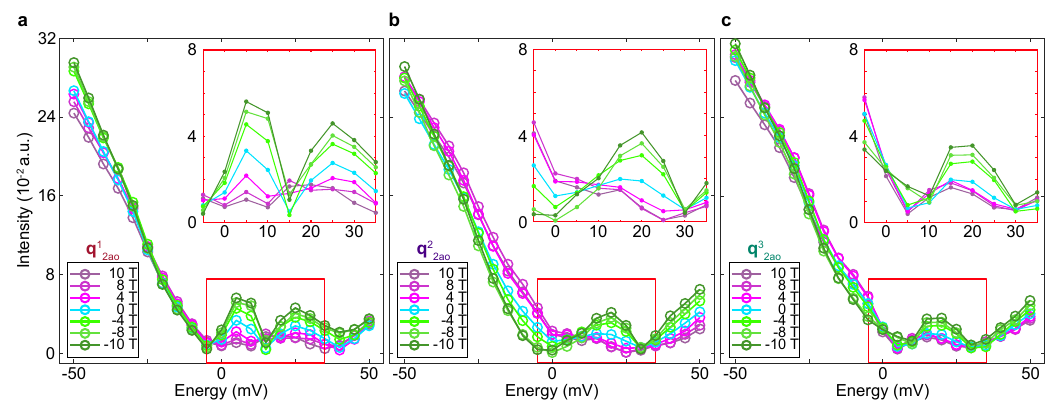}
    \caption{\textbf{Additional spin-polarized STM characterization at more magnetic fields.} - \textbf{a-c}, Intensities of the charge modulation peaks $\mathbf{q}_{2a0}^1$, $\mathbf{q}_{2a0}^2$, and $\mathbf{q}_{2a0}^3$ in d$I$/d$V$ maps as a function of energy for different magnetic fields over the same area as that in Figure 4 with the \textbf{q}-vectors defined in the same way. STM setup conditions: $V_{sample}$ = 50 mV, $I_{set}$ = 150 pA, $V_{exc}$ = 5 mV (rms).} 
    \label{suppfig:6}
\end{figure}

\end{document}